
\magnification=1200
\hoffset=-.1in
\voffset=-.2in
\vsize=7.5in
\hsize=5.6in
\tolerance 10000
\def\pmb#1{\setbox0=\hbox{$#1$}%
\kern-.025em\copy0\kern-\wd0
\kern.05em\copy0\kern-\wd0
\kern-.025em\raise.0433em\box0 }

\baselineskip 12pt plus 1pt minus 1pt
\pageno=0
\centerline{\bf POLARIZED $\pmb{q\to\Lambda}$ FRAGMENTATION}
\smallskip
\centerline{{\bf FUNCTIONS FROM  $\pmb{e^+ e^-\to\Lambda+X}$}\footnote{*}
{This work is supported in part by funds
provided by the U. S. Department of Energy (D.O.E.) under contract
\#DE-AC02-76ER03069, and the Texas National Research Laboratory Commission
under grant \#RGFY9266.}}
\vskip 24pt
\centerline{M.~Burkardt and R.~L.~Jaffe}
\vskip 12pt
\centerline{\it Center for Theoretical Physics}
\centerline{\it Laboratory for Nuclear Science}
\centerline{\it and Department of Physics}
\centerline{\it Massachusetts Institute of Technology}
\centerline{\it Cambridge, Massachusetts\ \ 02139\ \ \ U.S.A.}
\vskip 1.5in
\centerline{Submitted to: {\it Physical Review Letters}}
\vfill
\centerline{ Typeset in $\TeX$ by Roger L. Gilson}
\vskip -12pt
\noindent CTP\#2186\hfill February 1993
\eject
\baselineskip 24pt plus 2pt minus 2pt
\centerline{\bf ABSTRACT}
\medskip
Measurement of the helicity asymmetric cross section for
semi-inclusive production of $\Lambda$-hyperons in $e^+e^-$
annihilation near the $Z^0$ resonance allows a complete determination
of the spin-dependent fragmentation functions for the different quark
flavors into the $\Lambda$. The
parity violating, self analyzing,
decay of the final state $\Lambda$ makes the
experimental analysis of the helicity asymmetry possible.
This experiment should be practical with present day technology at the
LEP collider at CERN or at SLC at SLAC.
\vfill
\eject
\noindent{\bf I.\quad INTRODUCTION}
\medskip
\nobreak
In the symmetric
quark model, the $\Lambda$-baryon has a rather simple spin-flavor
wavefunction. All its spin is carried by the $s$-quark, while the $ud$-pair
is coupled to $S=0$, $I=0$. The $\Lambda$ thus seems to provide a particularly
``clean'' example to reexamine the spin-crisis,$^{1,\,2,\,3}$
In general one has
$$\int_0^1 dx\, g_1^{e\Lambda}(x) = {1\over2} \left(
{4\over 9} \Delta u^\Lambda
+ {1\over 9} \Delta d^\Lambda
+ {1\over 9} \Delta s^\Lambda  \right)\eqno(1)$$
where $\Delta q^\Lambda$ is the $\Lambda$ matrix element of the
$q$-quark axial charge or equivalently, the fraction of the
spin of the $\Lambda$ carried by the spin of quarks and antiquarks of
flavor $q$. The weak $Q^2$-dependence generated by QCD radiative corrections
has been ignored for simplicity.  In the nonrelativistic quark model
$\Delta u^\Lambda=\Delta d^\Lambda=0$ and $\Delta s^\Lambda=1$ so
$\int_0^1 dx\, g_1^{e\Lambda}(x) = {1/18}$. A more sophisticated
analysis makes use of information
coming from flavor $SU(3)$ octet axial charges from
hyperon $\beta$ decay:
$$\int_0^1 dx\, g_1^{e\Lambda}(x) =
{1\over18}\left( 2\Sigma -D \right)\eqno(2)$$
$F$ and $D$ are invariant matrix elements in $\beta$-decay, presently
estimated to be $F=0.47 \pm 0.04$ and $D=0.81\pm 0.03$, and $\Sigma$
is the flavor singlet quark spin operator. The assumption
$\left\langle N\left|\bar{s}\gamma_\mu \gamma_5s \right|N\right\rangle=0$$^1$
gives $\Sigma = 3F-D$ and predicts
$\int_0^1 dx\, g_1^{e\Lambda}(x) = 0.022 \pm 0.014$.
The most reliable prediction is obtained by using the proton data
to supply the necessary information on $\Sigma$,
$$\int_0^1 dx\, g_1^{e\Lambda}(x) = \int_0^1 dx\, g_1^{ep}(x)
- {1\over 18} \left(2D+3F\right)\ \ .\eqno(3)$$
This sum rule is on the same footing as the Bjorken sum rule$^4$
except that it relies on the full $SU(3)$-flavor symmetry
while Bjorken's requires only isospin invariance.
Using the EMC analysis
($\int_0^1 dx\, g_1^{ep}(x)=0.126\pm0.018$
we find    $\int_0^1 dx\, g_1^{e\Lambda}(x)=-0.042\pm0.019$, far from the
naive quark model or the $\left\langle
N\left|\bar{s}\gamma_\mu \gamma_5s \right|N\right\rangle=0$  prediction.

While the quark model identifies the $\Lambda$-spin with the spin of the
$s$-quark, the above analysis suggests that the actual situation
might be more complex. This becomes more obvious if one
uses $SU(3)$ to
decompose the last (most reliable) prediction (3) into
its $\Delta u$, $\Delta d$ and $\Delta s$ contributions, yielding
$$\eqalign{
\Delta u^\Lambda = \Delta d^\Lambda &= {1\over 3}(\Sigma - D)
= -0.23 \pm 0.06\cr
\Delta s^\Lambda &= {1\over 3}(\Sigma+2D) = +0.58 \pm 0.07\cr}\eqno(4)$$
as opposed to the naive expectation
$\Delta u^\Lambda=\Delta d^\Lambda=0$ and $\Delta s^\Lambda=1$.
Unfortunately, no $\Lambda$ targets are available for deep inelastic
scattering experiments and it thus seems impossible to actually
measure   $\int_0^1 dx\, g_1^{e\Lambda}(x)$.

In this Letter, we instead show how to to measure the polarized
fragmentation functions for the decay of quarks into a $\Lambda$
--- an experimental program that, as we will see below, seems to be realistic.
The program we describe requires measurement of total inclusive
$\Lambda$ production in $e^+e^-$ annihilation at various energies: off,
near and on the $Z^0$ peak. Provided $\Lambda$'s can be reconstructed
and their polarization measured in the usual fashion
through the self analyzing decay $\Lambda \rightarrow p \pi^-$,
the necessary fragmentation functions should be easy to measure.

There is a potentially important background from the process
$e^+e^-\to\Sigma^0+X$ followed by $\Sigma^0\to\Lambda\gamma$.  $\Lambda$'s
produced in this way are {\it not\/} part of the $q\to\Lambda$ fragmentation
function which includes strong interaction processes alone.  [Note,
$\Lambda$'s produced by strong decays of hyperon resonances, $Y^*\to\Lambda
X$, {\it are\/} properly included in the $\Lambda$ fragmentation function.]  A
precise experiment would be required to veto events in which a prompt photon
accompanies the produced $\Lambda$.

Fortunately, even though excluding secondary
$\Lambda$'s would help
to reduce systematic errors this is not crucial
since
the $\Sigma^0$ multiplicity in $e^+e^- \rightarrow$ hadrons
is typically about a factor 3.5 smaller$^5$ than
the $\Lambda$ multiplicity.   Furthermore,
the $\Lambda$'s from $\Sigma^0 \rightarrow \Lambda\gamma$
are depolarized by a factor ${1/3}$ compared to the initial
$\Sigma^0$'s, which is important here because we are only
interested in the helicity asymmetric cross section. Combining
the multiplicity suppression with the depolarization effect we arrive at
only a 10\% contamination for the helicity asymmetric
$\Lambda$ production cross section when secondary $\Lambda$'s
from $\Sigma^0$ decay are not vetoed.
\goodbreak
\bigskip
\noindent{\bf II.\quad FRAGMENTATION FUNCTIONS AND CROSS SECTIONS}
\medskip
\nobreak
In the parton model, the differential cross section for
$e^-e^+\rightarrow h+X$ is obtained by summing over the
cross sections for $e^+e^-\rightarrow q\bar{q}$, weighted with
the probability $d^h_q(z,Q^2)$ that a quark with momentum
${1\over z}P$ fragments into a hadron $h$ with momentum $P$$^6$
$${d^2\sigma^h\over d\Omega\, dz} = \sum_q {d\sigma^q\over d\Omega}
d^h_q(z,Q^2)\ \ .\eqno(5)$$
Here $q=k_{e^-} + k_{e^+}$, $Q^2=q^2=s>0$ and $z={2P\cdot q\over Q^2}$.
For a field theoretic definition of the
fragmentation function $d^h_q(z,Q^2)$ see for example Refs.~[7,8,9].
 The more sophisticated treatment
is equivalent to the parton model for our purposes so we use the
parton model language henceforth.
In the naive parton model the fragmentation functions depend
only on the scaling variable $z$. However, similar to deep inelastic
structure functions, fragmentation functions in QCD also show
logarithmic evolution with $Q^2$.$^{8,\,9}$
Energy conservation requires
$$\sum_h\int_0^1\,dz\, d^h_q(z,Q^2) z =1\ \ .\eqno(6)$$
In the following we concentrate on polarized fragmentation functions
$$\eqalignno{\Delta \hat{ q}(z) &=
d^{\Lambda(L)}_{q(L)}(z) - d^{\Lambda(R)}_{q(L)}(z)  \cr
&= d^{\Lambda(L)}_{q(L)}(z) - d^{\Lambda(L)}_{q(R)}(z) &(7) \cr
\Delta \hat{ \bar{q}}(z) &=
d^{\Lambda(L)}_{\bar{q}(L)}(z) - d^{\Lambda(R)}_{\bar{q}(L)}(z) \cr
 &=d^{\Lambda(L)}_{\bar{q}(L)}(z) - d^{\Lambda(L)}_{\bar{q}(R)}(z) &(8)\cr}$$
defined as the probability that a left-handed quark of flavor $q$
fragments
into a left handed $\Lambda$ (with momentum fraction $z$) minus
the probability that the left handed quark fragments into a
right-handed $\Lambda$. The interpretation of the antiquark
fragmentation function (8) is similar. For simplicity
we suppress the $Q^2$ dependence of the $d^h_q$.
In order to avoid confusion with similar observables in the context
of polarized deep inelastic scattering, we put a caret on all
fragmentation asymmetries.
For a measurement of these helicity asymmetric fragmentation functions
one needs to know both the polarization of the initial state (quark)
and the final state (baryon). In the case of the $\Lambda$-baryon
the final state polarization can be easily determined because
the (weak) decay $\Lambda \rightarrow \pi^-p$ violates parity.
In the rest frame of the $\Lambda$ the decay distribution of the
proton is$^{10}$
$$I(\theta) = {1\over 4\pi} \left(1 + a \cos \theta \right)\ \ ,\eqno(9)$$
($a= .642 \pm .013$$^5$
where $\theta$ is the angle between the momentum of the outgoing
proton (in the rest frame of the $\Lambda$) and the spin of the
$\Lambda$. For a more detailed discussion of this ``self-analyzing''
decay we refer to the literature.$^{10}$

We now consider $\Lambda$ production via photons and $Z^0$'s.
To exploit $e^-e^+$ annihilation via photons
one has to start from polarized $e^-$ (or $e^+$) in order to fix the
polarization of the quarks. Using (5), as well as the
shorthand notation for the asymmetries (7) and
(8), one thus finds for the helicity asymmetric cross-section
assuming $e^+e^- \rightarrow \gamma \rightarrow q\bar{q}$:
$$\eqalign{
{d^2\sigma^{e^-(L)e^+\rightarrow \Lambda(L)X}\over d\Omega \,dz}
&-{d^2\sigma^{e^-(L)e^+\rightarrow \Lambda(R)X}\over d\Omega \,dz}
={\alpha^2\over 2s}
\cos \theta
\sum_q Q_q^2 \left( \Delta \hat{ q}(z) + \Delta \hat{ \bar{q}}(z)\right) \cr
&={\alpha^2\over 2s} \cos \theta
\left[ {5\over 9}\left( \Delta \hat{ u}(z) + \Delta \hat{ \bar{u}}(z)\right)
+ {1\over 9}\left( \Delta \hat{ s}(z) + \Delta \hat{ \bar{s}}(z)\right)\right]
\cr}\eqno(10)$$
where $L,R$ denotes the helicity of the
$e^-$ and the $\Lambda$
($e^+$ unpolarized, polarization of $X$ not measured) and $\theta$
is the angle
between the momenta of the incoming $e^-$ and the outgoing $\Lambda$
in the CM frame ($\cos \theta=
\widehat{\vec{k}}_{e^-}\cdot\widehat{\vec{P}}_\Lambda$).
Here we have made use of isospin symmetry of the fragmentation functions
which implies for a $\Lambda$
$$\eqalignno{\Delta \hat{ d}(z) &= \Delta \hat{ u}(z)  &(11)\cr
\Delta \hat{ \bar{d}}(z) &= \Delta \hat{ \bar{u}}(z)  &(12)\cr}$$
[This is truly isospin symmetry,
as distinct from the ``isospin symmetry of the sea'' often discussed
in connection with the Gottfried Sum Rule].

At higher energies, where the $Z$-resonance
as well as $\gamma Z$ interference are relevant it is not necessary
to start from a polarized $e^-e^+$ state because the parity violating
coupling of the fermions favors certain helicity states.
In the standard electroweak theory, combined with parton model
assumptions, one obtains
$$\eqalign{
&{d^2\sigma^{e^-e^+\to\Lambda (L)X}\over d\Omega\,dz} - {d^2
\sigma^{e^-e^+\to\Lambda(R)X}\over d\Omega \,dz} = {\alpha^2\over 2s}\sum_q
\chi_1 \left( -Q_q\right) \bigl[ a_q v_e \left(\Delta \hat q(z) -
\Delta\hat{\bar{q}}(z)\right) \left( 1 + \cos^2\theta\right) \cr
&\quad +2a_e v_q \left( \Delta \hat q(z) + \Delta \hat{\bar{q}}(z)\right)
\cos\theta\bigr] + \chi_2 \bigl[ \left( v^2_e + a^2_e\right) v_q a_q \left(
\Delta \hat q(z) - \Delta \hat{\bar{q}}(z)\right) \left( 1 +
\cos^2\theta\right) \cr
&\quad +2v_e a_e
\left( v^2_q + a^2_q\right) \left( \Delta \hat q(z) + \Delta
\hat{\bar{q}}\right) \cos\theta] \cr}\eqno(13) $$
where$^5$
$$\eqalignno{ \chi_1 &= {1\over 16 \sin^2 \Theta_W \cos^2 \Theta_W}\
{s \left(s-M_Z^2\right)\over \left(s-M_Z^2\right)^2 +\Gamma_Z^2 M_Z^2}&(14)
\cr
\chi_2 &= {1\over 256 \sin^4 \Theta_W \cos^4 \Theta_W}\
{s^2\over \left(s-M_Z^2\right)^2 +\Gamma_Z^2 M_Z^2}\ \ .&(15)\cr}$$
$M_Z=91.17$~GeV and $\Gamma_Z=2.49$~GeV
are the mass and width of the $Z$. $v_e=4\sin^2\Theta_W-1$ and $a_e=-1$
are the vector and axial vector
couplings of the electron to the $Z$. Here we adopt the conventions of
the particle data group$^5$ where the coupling of a
fermion to the $Z$-boson is given by
$-{g\over 2\cos \Theta_W}
\bar{\psi} \gamma^\mu
\left( v - a\gamma_5 \right)\psi Z_\mu$. The
couplings of the quarks to
the $Z$ are $v_u=1-{8\over 3}\sin^2\Theta_W$,
$v_d=v_s=-1+{4\over3}\sin^2\Theta_W$, $a_u =1$ and $a_d=a_s=-1$.
Both $e^-$ and $e^+$ are unpolarized and the $L$, $R$ denotes the
helicity of the $\Lambda$.
Using again isospin symmetry for the fragmentation functions and inserting
the explicit expressions for the axial and vector couplings we find
$$\eqalign{
&{4s\over \alpha^2} \left[ {d^2\sigma^{e^-e^+\to\Lambda(L)X}\over
d\Omega\,dz} - {d^2\sigma^{e^-e^+ \to \Lambda(R)X}\over d\Omega\,dz}\right]\cr
&\quad = \chi_1 \biggl\{ \left[ c_1 \left( \Delta\hat u(z) -
\Delta\hat{\bar{u}}
(z)\right)
+ c_2 \left( \Delta \hat s (z) - \Delta \hat{\bar{s}}(z)\right)\right]
\left( 1 + cos^2\theta\right) \cr
&\qquad +  \left[ c_3 \left(\Delta\hat u(z) +
\Delta \hat{\bar{u}} (z)\right) + c_4
\left( \Delta \hat s(z) + \Delta \hat{\bar{s}}(z)\right)\right] \cos\theta
\biggr\} \cr
&\qquad +  \chi_2 \biggl\{
\left[ c_5\left( \Delta \hat u (z) - \Delta\hat{\bar{u}}(z)
\right) + c_6 \left( \Delta \hat s(z) - \Delta\hat{\bar{s}}(z)\right)\right]
\left( 1 + \cos^2\theta\right) \cr
&\qquad +
\left[ c_7 \left(\Delta\hat u (z) + \Delta \hat{\bar{u}}(z)\right) + c_8
\left( \Delta\hat s(z) + \Delta\hat{\bar{s}}(z)\right)\right] \cos\theta
\biggr\} \ \ .\cr}\eqno(16)$$
With $x_W = \sin^2\Theta_W= 0.2325 \pm 0.0008$,$^5$
the c's are given by
$c_1=-2v_e = 0.1400 \pm 0.0064$,
$c_2=-{2\over 3}v_e  = 0.0467 \pm 0.0022$,
$c_3=4\left(1-{20\over9}x_W\right) = 1.9333 \pm 0.0071$,
$c_4=4\left({1\over3}-{4\over9}x_W\right) = 0.9200 \pm 0.0014$,
$c_5=8\left(1-4x_W+8x^2_W\right)
\left(1-2x_W\right) = 2.1505 \pm 0.0074$,
$c_6=4\left(1-4x_W+8x^2_W\right)
\left(1-{4\over 3}x_W\right) = 1.3868 \pm 0.0028$,
$c_7=-16v_e\left(1-2x_W
+{20\over 9}x^2_W\right) = 0.7337 \pm 0.0343$,
$c_8=-8v_e\left(1-{4\over 3}x_W
+{8\over 9}x^2_W\right) = 0.4133 \pm 0.0193$.
\goodbreak
\bigskip
\noindent{\bf III.\quad DISCUSSION}
\medskip
\nobreak
In principle, Eqs.~(13) and (16) are sufficient
to determine all four independent fragmentation functions
($\Delta \hat{ u}(z)$, $\Delta \hat{ \bar{u}}(z)$,
$\Delta \hat{ s}(z)$ and $\Delta \hat{ \bar{s}}(z)$) of a $\Lambda$
separately. For example, the charge conjugation even combinations
$\Delta \hat{ q}(z) + \Delta \hat{ \bar{q}}(z)$ (which are proportional to
$\hat{G_1}(z)$) and the charge conjugation odd combination
$\Delta \hat{ q}(z) - \Delta \hat{ \bar{q}}(z)$ (which are proportional to
the parity violating fragmentation function
$\hat{X_1}(z)$$^{11}$ may be distinguished
by means of their behavior under $\theta \rightarrow \pi - \theta$.
The contribution from the three relevant quark flavors ($u=d$, $s$)
can be disentangled by varying the invariant mass and thus emphasizing
annihilation via photons or $Z$'s or the interference
term independently. However, there is one practical limitation
to this program: $c_1$, $c_2$, $c_7$ and $c_8$ are all proportional
to the vector-coupling of the $Z$ to an electron, $v_e = 4x_W-1$,
which is very small because $x_W$ is very close to
${1\over 4}$.   This does not limit the possibility of measuring
the  charge-even fragmentation functions because annihilation
via photons and the $\gamma Z$-interference term are sufficient for this
purpose. Furthermore, other numerical factors compensate for
the smallness of $\left(1-4x_W\right)$ in $c_7$ and $c_8$.
However, since charge-odd terms do not contribute to annihilation
via photons, the smallness of $c_1$ and $c_2$ may
restrict accurate measurements of the charge conjugation odd terms to
the linear combination
$$c_5\left(\Delta \hat{ u}(z) - \Delta \hat{ \bar{u}}(z)\right)
+c_6\left(
\Delta \hat{ s}(z) - \Delta \hat{ \bar{s}}(z)
\right)\ \ .\eqno(17)$$
Alternatively, one can start from polarized
$e^-$ for energies near the $Z$ resonance, where one finds
$$\eqalign{
&{4s\over \alpha^2} \left[ {d^2\sigma^{e^-(L)e^+\to\Lambda(L)X}\over
d\Omega\,dz} - {d^2\sigma^{e^-(L)\to\Lambda(R)X}\over d\Omega\,dz}\right]
=2\sum_q Q^2_q \left(\Delta \hat q(z) + \Delta\hat{\bar{q}} (z)\right)
\cos\theta \cr
&\quad + \chi_1 \left( - Q_q\right) \left( v_e + a_e\right) \left[ a_q \left(
\Delta\hat q(z) - \Delta\hat{\bar{q}} (z)\right) \left( 1 +
\cos^2\theta\right)
+ 2v_q \left( \Delta\hat q(z) + \Delta \hat{\bar{q}}(z)\right)
\cos\theta\right] \cr
&\quad + \chi_2 \left( v_e + a_e\right)^2 \left[ v_q a_q \left(\Delta\hat q(z)
-
\Delta\hat{\bar{q}}(z)\right) \left( 1 + \cos^2\theta\right) + \left( v^2_q +
a^2_q\right) \left( \Delta\hat q (z) + \Delta \hat{\bar{q}}(z)\right)
\cos\theta\right] \cr
&= 2\left[ {5\over 9} \left(
\Delta \hat u(z) + \Delta\hat{\bar{u}}(z)\right) +
{1\over 9} \left(\Delta\hat s(z) +\Delta \hat{\bar{s}} (z)\right)\right]
\cos\theta \cr
&\quad +  \chi_1\bigl\{ \left[\tilde c_1 \left( \Delta \hat u (z) -
\Delta\hat{\bar{u}}(z)\right) + \tilde c_2 \left(\Delta\hat s(z) -
\Delta\hat{\bar{s}}(z)\right)\right] \left( 1 + \cos^2\theta\right)\cr
&\quad +\left[
\tilde c_3 \left( \Delta\hat u(z) + \Delta\hat{\bar{u}} (z)\right) + \tilde
c_4 \left(\Delta\hat s(z) + \Delta\hat{\bar{s}}(z) \right)\right]\cos \bigr\}
\cr
&+  \chi_2 \bigl\{ \left[ \tilde c_5 \left( \Delta\hat u(z) -
\Delta\hat{\bar{u}} (z)\right) + \tilde c_6 \left(\Delta \hat s (z) - \Delta
\hat{\bar{s}}(z) \right)\right] \left( 1 + \cos^2\theta\right)\cr
&\quad  + \left[ \tilde
c_7 \left(\Delta\hat u (z) + \Delta\hat{\bar{u}}(z)\right) + \tilde c_8 \left(
\Delta \hat s (z) + \Delta \hat{\bar{s}}(z)\right)\right]\cos\theta\bigr\}\ \
,\cr}\eqno(18)$$
where
$\tilde{c}_1=4\left( 1-2x_W\right)= 2.1400 \pm 0.0064$,
$\tilde{c}_2= {1\over 3}\tilde{c}_1 = 0.7133 \pm 0.0021$,
$\tilde{c}_3= 2\tilde{c}_1
\left(1-{20\over 9}x_W\right) = 2.0687 \pm 0.0138$,
$\tilde{c}_4=2\tilde{c}_1
\left({1\over 3}-{4\over 9}x_W\right) = 0.9844 \pm 0.0045$,
$\tilde{c}_5={1\over 4}\tilde{c}_1^3 = 2.4501 \pm 0.0119$,
$\tilde{c}_6={1\over 2}\tilde{c}_1^2
\left(1-{4\over 3}x_W\right) = 1.5800 \pm 0.0119$,
$\tilde{c}_7= 2\tilde{c}_1^2 \left(1-2x_W
+{20\over 9}x^2_W\right) = 6.0004 \pm 0.0428$,
$\tilde{c}_8=  \tilde{c}_1^2
\left(1-{4\over 3}x_W
+{8\over 9}x^2_W\right) = 3.3800 \pm 0.0236$.
Note that $\tilde{c}_1$ and $\tilde{c}_2$ are not small compared
to $\tilde{c}_3$ and $\tilde{c}_4$
making it easier to
extract the charge odd asymmetries for $u$ and $s$ quarks separately.
As far as disentangling the contributions from the various quark
flavors is concerned, the situation is better in annihilation
than in deep inelastic
scattering off nucleons. The fragmentation into $\Lambda$'s
allows the measurement of 4 linearly independent,  spin-dependent,
observables at leading twist
(actually 6, if one makes use of isospin symmetry).
Equivalent measurements in deep inelastic scattering off nucleons
would require the combinations of electromagnetic as well as charged
current data from polarized protons and neutrons --- a very difficult
challenge.

Unfortunately, very little is known theoretically about fragmentation
functions --- especially the helicity-odd fragmentation functions discussed
here. For example, there are no sum-rules known, since the
moments of fragmentation functions are not related
to hadron expectation values of local
operators.
Also there is little guidance from theory on the z-dependence
of the polarized fragmentation functions.
In a naive quark model for
the $\Lambda$, one would expect $\Delta \hat{ s}(z)$ to be positive,
while all other fragmentation functions ($\Delta \hat{ \bar{s}}(z)$,
$\Delta \hat{ u}(z)$, $\Delta \hat{ \bar{u}}(z)$) should vanish.
However, our experience with polarized deep inelastic structure functions
suggests that this picture will most likely be modified.
The existence of such a straightforward experimental program
to measure the flavor dependence of polarized fragmentation
functions should spur the theoretical community to consider these
quantities.
\goodbreak
\bigskip
\centerline{\bf ACKNOWLEDGEMENTS}
\medskip
\centerline{It is a pleasure to thank Xiangdong Ji
for many discussions.}
\vfill
\eject
\centerline{\bf REFERENCES}
\medskip
\item{1.}
J.~Ellis and R.~L.~Jaffe, {\it Phys.~Rev.\/} {\bf D9}, 1445 (1974).
\medskip
\item{2.}
J.~Ashman {it et al.\/}, {\it Phys.~Lett.\/} {\bf B206}, 364 (1988);
{\it Nucl.~Phys.\/} {\bf B328}, 1 (1989).
\medskip
\item{3.}M.~J.~Alguard {\it et al.\/}, {\it Phys.~Rev.~Lett.\/}
 {\bf 37}, 1261 (1976);
{\bf 41}, 70 (1978); G.~Baum {\it et al.\/}, {\it ibid.} {\bf 51}, 1135 (1983).
\medskip
\item{4.}J.~D.~Bjorken, {\it Phys.~Rev.\/} {\bf 148}, 1467 (1966).
\medskip
\item{5.}Particle Data Group, {\it Phys.~Rev.\/} {\bf D45}, 1 (1992).
\medskip
\item{6.}R.~D.~Field, {\it Applications of Perturbative QCD\/} (Addison-Wesley,
Redwood City, 1989).
\medskip
\item{7.}R.~L.~Jaffe and X.~Ji, in preparation.
\medskip
\item{8.}J.~C.~Collins and D.~E.~Soper, {\it Nucl.~Phys.\/}
{\bf B194}, 445 (1982).
\medskip
\item{9.}I.~I.~Balitsky and V.~M.~Braun, {\it Nucl.~Phys.\/}
{\bf B361}, 93 (1991).
\medskip
\item{10.}D.~H.~Perkins, {\it Introduction to High Energy Physics\/}
 (Addison-Wesley, Reading, MA, 1982).
\medskip
\item{11.}X.~Ji, submitted to {\it Nucl.~Phys.~B}, CTP 2141.
\par
\vfill
\end